\documentclass{edp-jp4}
\usepackage{graphicx}
\usepackage{amsmath}
\begin{document}

\title{Coulombian disorder in Charge Density Waves} 
 \author{Alberto Rosso}\address{ Laboratoire de physique th\'{e}orique et
mod\`{e}les statistiques CNRS UMR8626 \\ B\^{a}t. 100 Universit\'{e}
Paris-Sud; 91405 Orsay Cedex, France; \email{rosso@lptms.u-psud.fr}}
\author{Edmond Orignac} \address{Laboratoire de physique th\'eorique de
l'\'Ecole Normale Sup{\'{e}}rieure, \\ CNRS UMR8549, 24, Rue Lhomond, 75231
Paris Cedex 05, France; \email{orignac@lpt.ens.fr}}
\author{R. Chitra}\address{Laboratoire de physique th\'eorique de la mati\`ere
condens\'ee CNRS UMR7600\\ Universit\'e de Pierre et Marie Curie, 4, Place
Jussieu 75252 Paris Cedex 05 France; \email{chitra@lptl.jussieu.fr}}
\author{Thierry Giamarchi}\address{Universit\'e de Gen\`eve, DPMC, 24 Quai
Ernest Ansermet, CH-1211 Gen\`eve 4, Switzerland;
\email{giamarchi@physics.unige.ch}}
\maketitle
\begin{abstract}
  When unscreened charged impurities are introduced in charge density wave
  system the long wavelength component of the disorder is long ranged and
  dominates static correlation functions. We calculate the x-ray intensity and
  find that it is identical to the one produced by thermal fluctuations in a
  disorder-free smectic-A. We suggest that this effect could be observed in
  the blue bronze material K$_{0.3}$MoO$_3$ doped with charged impurities such
  as Vanadium.
\end{abstract}
\section{Introduction}

Quasi-one dimensional material formed of weakly coupled chains are known to
posses low temperature instability towards a state with an incommensurate
static modulation of the electronic charge and of the lattice displacement
known as charge density wave (CDW) phase. CDWs behave as elastic systems
\cite{gruner_book_cdw,nattermann_brazovskii} provided Coulomb interaction are
screened. When Coulomb interaction are not sufficiently screened elasticity
becomes non-local \cite{efetov77_cdw,lee_coulomb_cdw} and disorder becomes
long ranged correlated. An important question is whether the competition of
the non-local elasticity with the long-range random potential results in the
complete destruction of the crystalline order, or whether some remnants of the
underlying periodic structure remain observable. A powerful experimental
technique able to characterize the structural properties of CDW with
impurities is afforded by x-ray diffraction. In this paper we derive the
static displacement correlation functions and x-ray intensity of the CDW with
charged impurities and we highlight the similitude of the latter to the x-ray
intensity of smectic-A liquid crystals subjected to thermal
fluctuations. Finally, we discuss the experimental significance of our result
and suggest that the smectic-like correlations should be observable in
experiments on $\mathrm{KMo_{1-x}V_xO_3}$.

\section{Charge density wave with charged impurities}
 We consider an incommensurate CDW, in which the electron density is modulated
by a wave vector $Q$ incommensurate with the underlying crystal
lattice. In this phase, the electron density has the following
form~\cite{gruner_book_cdw,nattermann_brazovskii}:
\begin{equation}
 \rho({\bf r}) = \rho_0 +\frac{\rho_0}{Q^2} {\bf Q}\cdot
 \nabla\phi({\bf r}) 
 +\rho_1 \cos( {\bf Q}\cdot {\bf r} +\phi({\bf r}))) \,,
\label{eq:rhoexp}
\end{equation}
where $\rho_0$, is the average electronic density. The second term in
Eq.~(\ref{eq:rhoexp}) is the long wavelength density  
and corresponds to variations of the density  over scales larger than $
Q^{-1}$ . The last oscillating term describes the sinusoidal deformation of the
density at a scale of the order of $Q^{-1}$ induced by the formation
of the CDW with amplitude $\rho_1$  and phase $\phi$. 
 
The low energy properties of the CDW are described by an elastic Hamiltonian
for the fluctuations of the phase $\phi$. When the Coulomb interactions are
unscreened this Hamiltonian becomes non-local. Choosing the $x$ axis aligned
with ${\bf Q}$ the total Hamiltonian in three dimensions reads
\cite{efetov77_cdw,lee_coulomb_cdw}:
 
\begin{eqnarray} 
\label{eq:anisotropyfinal} 
H_{\mathrm{el.}}&=&H_0+H_C=\frac 1 2 \int G^{-1}(q) |\phi(q)|^2,  \\
 G^{-1}(q)&=& \frac{n_c \hbar v_F v_\perp^2 }{2\pi v_x^2}
 \left[\frac{q_x^2}{{\bf q}^2\xi_0^2} +  {\bf q}_\perp^2 \right]  \nonumber\
\end{eqnarray}
\noindent
 where $v_F$ is the Fermi velocity and $n_c$ is the number of chains per unit
 surface that crosses a plane orthogonal to $Q$.  The velocity of the phason
 excitations parallel to $Q$ is $v_x=(m_e/m_*)^{1/2} v_F$ with $m^*$ the
 effective mass of the CDW and $m_e$ the mass of an electron.  $v_\perp$
 denote the phason velocities in the transverse directions. The lengthscale
 $\xi_0$ is defined by:
 \begin{eqnarray}
   \label{eq:xi-definition}
   \xi_0^2 = \frac{n_c \hbar v_F v_\perp^2 }{2\pi v_x^2 e^2 \rho_0^2}
   Q^2\epsilon.
 \end{eqnarray}
\noindent
where we have assumed for simplicity an isotropic dielectric permittivity
$\epsilon$ of the host medium and $e$ is the charge of the electron. 
When a non-zero concentration of charged
impurities is present, the Hamiltonian becomes: 
\begin{eqnarray}
\label{eq:Hamiltonian2}
H&=& H_{\mathrm{el.}}
+\frac{e^2}{4\pi \epsilon}  \int d^3{\bf r} d^3{\bf r'} 
     \frac{\rho_{\mathrm{imp.}}({\bf r}) \rho({\bf r'})} {|r-r'|}, \nonumber \\
\end{eqnarray}
\noindent where $\rho_{imp}(\textbf{r})$ is the density of impurities at
position ${\bf r}$. We assume Gaussian distributed impurities with
$\overline{\rho_{imp}(\textbf{r}) \rho_{imp}(\textbf{r'})} = D
\delta(\textbf{r} - \textbf{r'})$, where $D$ is the measure of the disorder
strength.  Using the expression of the density in~(\ref{eq:rhoexp}), the
interaction with disorder comprises two parts, one that describes the
interaction with long wavelength fluctuations of the density (forward
scattering), and a second part that describes interactions of impurities with
density fluctuations of wavelength $Q^{-1}$ (backward scattering). In
\cite{orignac_cdw} we have shown that even in the case of long range disorder,
the backward scattering terms behave essentially like their short ranged
(neutral impurities) counterparts. In that case, it is known that in three
dimensions, the contribution of the backward scattering to the phase
correlations grows slowly as $\log (\log
r)$.\cite{chitra_vortex,rosso_cdw_long} The contribution of the forward
scattering term is obtained from the Hamiltonian:
\begin{eqnarray}
\label{eq:Hamiltonian3}
H = \int \frac  {d^3q}{(2\pi)^3}\left[ \frac{G^{-1}(q)}{2} |\phi(q)|^2 +
  \frac{i\rho_0 e^2 q_x}{Q\epsilon q^2 } 
  \rho_{\text{imp.}}(-q) \phi(q) \right], \nonumber \\     
\end{eqnarray}
and reads in the limit of $r_\perp,x \to \infty$:
\begin{eqnarray}
\label{eq:correlator}
B(r)&=&\overline{\langle(\phi(r)-\phi(0))^2\rangle} \,\nonumber \\
 &=& \kappa   \ln \left(\frac{r_\perp^2
  +4|x|\xi_0}{\Lambda_{\perp}^{-2}}\right).
\end{eqnarray}
\noindent where $\Lambda_\perp$ is a momentum cutoff and: 
\begin{eqnarray}\label{eq:kappa-def}
  \kappa=\frac{D Q^2}{16\pi \xi_0 \rho_0^2}=\frac{D Q e v_x}{8 \rho_0 v_\perp}
  \frac{1}{\sqrt{n_c h v_F \epsilon}}.
\end{eqnarray}
As a result, in the
presence of charged impurities, the forward scattering
terms generate the leading contribution to the phase correlations. 
\section{x-Ray spectrum and analogy with smectics A}

It is well known that the presence of a CDW in a compound is
associated with  the appearance of two satellites at positions
$q\sim K\pm Q$ around each
Bragg peak.\cite{cowley_x-ray_cdw} The intensity profiles of these
satellites give access to the structural properties of the CDW. 
 The main contribution to the x-ray satellite intensity is given by
$I_{\text{d}}$, which is symmetric under inversion around the Bragg
  vector $K$.\cite{ravy_x-ray_whiteline,brazovskii_x-ray_cdwT,rouziere_friedel_cdw,rosso_cdw_short,rosso_cdw_long}
The intensity $I_d$ is given by the following correlation function:
\begin{eqnarray}
  \label{eq:sym-simplified}
   I_{\text{d}}({\bf K + Q + k})  &=&  u_0^2 \overline{f}^2 K^2 \int d^3{\bf
 r}\, e^{-i \mathbf{k\cdot r} }
 C_{\text{d}}(\mathbf{r}) \,, \nonumber \\
 C_{\text{d}}(\mathbf{r})  &=& \left\langle \overline
 {e^{i(\phi(\mathbf{r}/2)-\phi(-\mathbf{r}/2))}} \right \rangle \,
\end{eqnarray}
\noindent where $\overline{f}$ is an average x-ray scattering amplitude and
$u_0$ is the amplitude of the lattice modulation induced by the presence of
the CDW.  Using the result given in Eq.(\ref{eq:correlator}) we extract the
small $k$ behavior of the intensity peak:
\begin{equation}
 I_{\text{d}}({\bf K + Q + k}) \sim
\begin{cases}
 (|k_x|)^{\kappa-2}&  \text{for $k_{\perp}^2 \xi_0 \ll |k_x|$,} \\
(|k_{\perp}|)^{2(\kappa-2)}& \text{otherwise.}
\end{cases}
\end{equation}
The intensity $I_d({\bf K + Q + k})$ is
divergent for $\kappa<2$ but is finite for $\kappa>2$, i.e. for strong
disorder.

We note that these intensities are remarkably similar to
those of a disorder-free smectic-A liquid
crystal\cite{caille_smectic_xray} 
at  positive temperature.  In fact, the expression of the exponent
$\kappa$ Eq.~(\ref{eq:kappa-def}) is analogous to the expression
(5.3.12) in~\cite{chandrasekhar_smectic}, with  the disorder
strength $D$ playing the role of the temperature $T$ in the smectic-A liquid
crystal. 

 Let us turn to an estimate of the exponent $\kappa$ appearing in the
intensities to determine whether such smectic-like intensities are indeed
observable in experiments.  A good candidate is the blue bronze material
$\mathrm{K_{0.3}MoO_3}$ which has a full gap so that Coulomb interactions are
unscreened at low temperature. We use the parameters of \cite{pouget_bronzes}:
\begin{eqnarray}
  n_c=10^{20}\text{chains}/m^2,\\
  v_F=1.3\times 10^5 m.s^{-1}, \\ 
  \rho_0=3 \times 10^{27} e^-/m^3, \\ 
  Q=6\times 10^9 m^{-1},
\end{eqnarray}
and a relative permittivity of $\epsilon_{\mathrm{K_{0.3}MoO_3}}=1$, so that
$\epsilon$ is equal to the permittivity of the vacuum.  For the doped material
$\mathrm{K_{0.3} Mo_{1-x} V_x O_3}$, we find that the disorder strength can be
expressed as a function of the doping and obtain:
\begin{equation}\label{eq:banal}
  D=x(1-x)\frac{\#(\text{Mo atoms/unit cell})}{\mathbf{a}\cdot(\mathbf{b}\times\mathbf{c})}.
\end{equation}
For the crystal parameters, $a=$18.25\AA, $b=$7.56\AA, $c=$9.86\AA,
$\beta=$117.53$^o$ \cite{sato_xray_kmoo3}, with 20 Molybdenum atoms per unit
cell, and a doping $x=3\%$, the disorder strength $D=4.8\times 10^{26}
\mathrm{m}^{-3}$.  Moreover, using the experimental bounds of the velocities:
$3.6\times 10^2 m.s^{-1}<v_\perp<1.6\times 10^3 m.s^{-1}$ and $v_x=3.7\times
10^3 m.s^{-1}$, we find that $\kappa$ is in the range $[0.16-0.8]$. Therefore,
the smectic-like order should be observable in x-ray diffraction measurements
on this material.


\end{document}